\documentstyle[epsfig]{mn}

\begin{document}

\title[The evolution of DDRGs]{Radio galaxies with a
`double-double' morphology:\\
{\Large \bf II - The evolution of double-double radio galaxies and implications
for the alignment effect in FRII sources}}
\author[C.R. Kaiser, A.P. Schoenmakers \& H.J.A. R{\"o}ttgering]{Christian R. Kaiser$^{1,2}$\thanks{email: ckaiser@mpa-garching.mpg.de}, Arno P. Schoenmakers$^{3,4,5}$ and Huub J. A. R{\"o}ttgering$^{4}$\\
$^1$ University of Oxford, Department of Physics, Nuclear Physics
Laboratory, Keble Road, Oxford, OX1 3RH, UK\\
$^2$ Max-Planck-Institut f{\"u}r Astrophysik,
Karl-Schwarzschild-Str. 1, 85740 Garching bei M{\"u}nchen, Germany\\
$^3$ Astronomical Institute, Utrecht University, P.O. Box 80 000, 3508 TA Utrecht, The Netherlands\\
$^4$ Sterrewacht Leiden, Leiden University, P.O. Box 9513, 2300 RA
Leiden, The Netherlands\\
$^5$ NFRA, P.O. Box 2, 7990~AA Dwingeloo, The Netherlands}

\maketitle

\begin{abstract}
A Double-Double Radio Galaxy (DDRG) is defined as consisting of a pair
of double radio sources with a common centre. In this paper we present
an analytical model in which the peculiar radio structure of DDRGs is
caused by an interruption of the jet flow in the central AGN. The new
jets emerging from the restarted AGN give rise to an inner source
structure within the region of the old, outer cocoon. Standard models
of the evolution of FRII sources predict gas densities within the
region of the old cocoon that are insufficient to explain the observed
properties of the inner source structure. Therefore, additional
material must have passed from the environment of the source through
the bow shock surrounding the outer source structure into the
cocoon. We propose that this material is warm clouds ($\sim\!10^4$ K)
of gas embedded in the hot IGM which are eventually dispersed over the
cocoon volume by surface instabilities induced by the passage of
cocoon material. The derived lower limits for the volume filling
factors of these clouds are in good agreement with results obtained
from optical observations. The long time scales for the dispersion of
the clouds ($\sim\!10^7$ yr) are consistent with the apparently
exclusive occurrence of the DDRG phenomenon in large ($\ga 700$ kpc)
radio sources and with the observed correlation of the strength of the
optical/UV alignment effect in $z\!\sim\!1$ FRII sources with their
linear size.
\end{abstract}

\begin{keywords}
galaxies: active -- galaxies: evolution -- galaxies: jets --
intergalactic medium
\end{keywords}

\section{Introduction}
\label{sec:intro}

Double-double radio galaxies (DDRGs) are defined as consisting of two
unequally sized, two-sided, double-lobed, edge-brightened (type FRII)
radio sources (see also the preceding paper Schoenmakers et
al. 1999a\nocite{sbrlk99}, hereafter paper I). The radio cores of the
two structures coincide. This gives them the appearance of two
independent FRII-type sources the smaller of which is placed inside
the larger structure. This morphology distinguishes the DDRGs clearly
from `normal' FRII objects. Seven DDRGs in which the inner and outer
structures are well aligned have been presented in paper I. In all
these objects the radio axes of the two structures form angles of less
than 7$^{\circ}$. The projected linear sizes of the outer source
structure of all these objects is greater than 700 kpc. Although it is
too early for final conclusions, we believe that large linear sizes of
the outer source structure is an intrinsic property of the DDRGs as a
sub-class of the FRII radio source population and not a selection
effect. The so-called X-shape radio sources (e.g. Leahy \& Williams
1984\nocite{lw84}) may be further examples of DDRGs but they differ
from the sources presented in paper I in that the radio axes of the
two structures are not aligned.

In most of the outer lobes of the DDRGs presented in paper I no radio
hot spots are detected.  For the exception to this rule, B\,1834+620,
only one hot spot is detected (Schoenmakers et
al. 1999b\nocite{sbrl99}, hereafter paper III). This may indicate that
the outer source structure of DDRGs is no longer supplied with energy
from the AGN via jets.

The main aim of this paper is to show that the observed properties of
the DDRGs are consistent with a model in which the jet production
mechanism in the AGN in the centre of the host galaxy of these objects
is interrupted and then restarted. During the first phase of activity
the outer radio structure is inflated while after the disruption the
inner structure is created. The formation of the inner structure of
the aligned DDRGs is only possible if some material from the outside
of the outer cocoon has penetrated the cocoon boundary and forms a
rather smooth density distribution within the region of the outer
cocoon. We show that this material is most likely provided by the
remains of warm, dense clouds which form one phase in the Inter
Galactic Medium (IGM) surrounding the DDRGs and are slowly dispersed
by the effects induced by the cocoon material streaming along their
surface.

In Sect \ref{sec:enviro} we briefly review the properties of the
gaseous environment of FRII radio sources. Section \ref{sec:disr}
discusses a possible mechanism for the restarting of the jet flow and
its implications for X-shaped radio sources. A brief review of the
analytical model used for analyzing the radio properties of the DDRGs
is given in Sect. \ref{sec:evo}. This model is then applied to the
outer, Sect. \ref{sec:old}, and the inner source structure, Sect.
\ref{sec:new}. In this Sect. we also show that some material from
the outside of the outer cocoon must have penetrated into this
region. Section \ref{sec:cont} reviews possible mechanisms for
achieving this contamination. 
Several implications that this contamination of the cocoon has for observable properties of the DDRGs will be discussed
in Sect. \ref{sec:impl}.
The optical and UV continuum emission of
the hosts of FRII sources at high redshift ($z> 0.6$) has been found
to be aligned with the radio source axis (for a review see McCarthy
1993\nocite{pm93}). In Sect. \ref{sec:opt} the implications for the
alignment effect in FRII sources of the most likely contamination
process, the shredding by the bow shock of warm, dense clouds embedded
in the otherwise hot IGM is investigated.

Throughout this paper we assume $H_o = 50$ km s$^{-1}$ Mpc$^{-1}$ and
$q_o=0.5$.

\section{The environments of FRIIs}
\label{sec:enviro}

The radio structures of radio galaxies are embedded in the Inter
Stellar Medium (ISM) of their host galaxies and, in the case of large
linear sizes, the IGM in between galaxies. 
Extended optical line emission observed to be associated with the
hosts of powerful radio galaxies at low redshift implies the existence
of warm gas ($T_{cl} \sim 10^4$ K) on scales of 10 kpc. Baum \&
Heckman (1989)\nocite{bh89} find that the density of this gas is at
least 0.1 cm$^{-3}$. This assumes a volume filling factor for this
material, $f_{cl}$, close to unity. This density is a lower limit
since for a `clumpy' distribution of the line emitting material,
i.e. smaller filling factors, higher densities are required. The gas
densities and the corresponding filling factors of the gas clouds
providing the extended optical emission observed in FRIIs are
difficult to constrain because of the poorly constrained properties of
the ionizing radiation originating in the central AGN and/or shocks in
the hot IGM which illuminates the gas in radio galaxies. The gas
density can in some cases be directly inferred from the flux ratios of
density-dependent emission lines, e.g. [S{\scshape ii}] 6717,
6732. For example, van Breugel et al. (1985)\nocite{bmhbb85} find gas
densities of $\sim 300$ cm$^{-3}$ for the line-emitting clouds in the
environment of the radio source 3C 277.3 and Heckman et
al. (1989)\nocite{hbbm89} find comparable densities for the emission
line regions of galaxies at low redshift associated with cooling
flows.

At high redshift ($z \ga 0.6$) the extended optical and UV emission,
line and continuum, of the host galaxies of FRII radio sources is
often aligned with the axis of the radio structure (for a review see
McCarthy 1993\nocite{pm93}). These structures can extend over several
100 kpc. Assuming that the aligned optical line emission in radio-loud
quasars is caused by the ionisation of warm gas by the central AGN,
the ionizing radiation of which can be to some extent constrained from
X-ray and UV observations, Heckman et al. (1991)\nocite{hlmb91} find
densities for the warm material of typically 100 cm$^{-3}$ while the
volume filling factors, $f_{cl}$, are estimated at $10^{-8} \rightarrow
10^{-7}$. These low values imply the existence of warm, dense clouds
embedded and in pressure equilibrium with the hot, X-ray emitting
phase of the IGM. The existence of such clouds also in low redshift
radio galaxies is consistent with the observations (Baum \& Heckman
1989\nocite{bh89}), if the volume filling factor of these clouds is
roughly 10$^{-3}$ on scales of 10 kpc and lower on larger scales in
these objects. Note however, the different distances from the center
of the host galaxy at which the line emission is observed in objects
at low and at high redshifts. It is not clear whether the properties
of the environments of FRIIs at low and at high redshift are similar
in this way.

Since all currently known DDRGs are large radio sources with linear
sizes $\ga 700$ kpc, the properties of the IGM rather than those of
the ISM determine the evolution and appearance of the outer source
structures of these objects. The density of the hot IGM ($T_x \sim
10^7$ K) at distances on a scale of 100 kpc can in principle be
inferred from its thermal bremsstrahlung emission in X-rays. On the
properties of the IGM on scales of Mpc, which is comparable to the
size of the outer structures of the DDRGs, we have no observational
constraints in these sources and we therefore have to estimate these
quantities. FRII radio sources at low redshift are often found in poor
groups of galaxies (e.g. Prestage \& Peacock 1988\nocite{pp88}) and
X-ray observations suggest that the density distribution of the gas in
such environments is well described by a King (1972)\nocite{ik72}
profile with $n_o = 10^{-2}$ cm$^{-3}$, $a_o =10$ kpc and $\beta
_{King} =0.5$ (Willott et al. 1999\nocite{wrbl99}, based on X-ray data
presented by Mulchaey \& Zabludoff 1998\nocite{mz98}). We will assume
in the following sections that the outer lobes of the DDRGs are
embedded in a hot IGM following such a density profile.

The observation of warm gas at large distances from the host galaxies
of radio sources implies that the surrounding IGM is a two-phase
medium. For the evolution of the large scale radio structure of FRIIs
only the hot ($T_x \sim 10^7$ K) phase is important. This can be seen
as follows.

The expansion of the large scale structure of FRII radio galaxies is
confined by the ram pressure of the material surrounding it. The warm
clouds will therefore be dynamically unimportant for the expansion of
the bow shock and cocoon of FRII sources as long as their volume
filling factor in the IGM is smaller than $\left( n_x / n_{cl} \right)
^{1/2}$, where $n_x$ is the density of the hot phase of the IGM and
$n_{cl}$ is the density of the warm clouds (Begelman \& Cioffi
1989\nocite{bc89}). For the warm clouds to be dynamically stable they
must be in pressure equilibrium with the hot gas and from this, for
the assumption of ideal gas conditions, we find $f_{cl} < \left(
T_{cl} / T_x \right) ^{1/2} \sim 0.03$. This will hold in virtually
every radio galaxy (but see for example McCarthy, van Breugel \&
Kapahi 1991\nocite{mbk91}). The expansion velocity of the outer cocoon
in DDRGs is therefore determined by the density distribution of the
hot IGM alone.

\section{The evolution of DDRGs}
\label{sec:over}

In this section we present a model for the evolution of the peculiar
radio structures of DDRGs. This evolution is crucially influenced by
the properties of the source environment outlined above. In this model
we take the double-double appearance of the DDRGs and the absence of
hot spots in most of the outer cocoons as evidence that the jet
activity in these sources first inflates the outer or `old' cocoon, is
then stopped by some mechanism disrupting the jet production process
in the central AGN and finally restarts causing the formation of the
inner or `new' cocoon.

Of the seven aligned DDRGs presented in paper I we exclude 3C 445 and
3C 219 from the analysis described in the following. In the case of 3C
445 sufficiently accurate radio flux measurements of the inner
structure are not available. The exceptionally high asymmetry of the
inner source structure of 3C 219 sets it apart from the other sources
considered here. This suggests that a process different from the one
described below is causing the formation of the double-double
structure in this object (see also paper I).

\subsection{Restarting of the jet flow}
\label{sec:disr}

The high degree of symmetry of the inner source structures of the
remaining five aligned DDRGs about the cores of the sources suggests
that the jet flow must have been restarted on both sides of the
sources simultaneously. The conclusions derived from the model
presented here do not depend on the exact mechanism(s) that cause the
disruption of the old jets and we are therefore unable to constrain
them. For a brief discussion of the possibilities we refer the reader
to paper I. Here we only present some further remarks on one possible
scenario, the recent infall of a large amount of gas onto the AGN.

If the restarting of the jets is caused by the infall of a large mass
of gas onto the AGN, there is no reason why the angular momentum
vector of the new material should have a direction similar to that of
the material in the original disk defining the direction of jet
propagation. This may imply that in general the direction of the new
jets deviates considerably from the old radio axis. In this case the
new jets will quickly propagate through a part of the old cocoon and
then inflate a new cocoon within the same environment as the old
jets. The radio emission of the old cocoon will fade quickly as will
be described in Sect. \ref{sec:old} but for some time four radio
lobes will be observable; two with hot spots at their ends and two
without. This is reminiscent of the so-called X-shaped or winged radio
galaxies (e.g. Leahy \& Williams 1984\nocite{lw84}). In this scenario
the aligned DDRGs of paper I are simply those sources in which the
angular momentum vector of the infalling material is not very
different from that of the existing accretion disk. Dennett-Thorpe et
al. (1998)\nocite{dblpr98} note that winged radio sources have radio
luminosities close to the FRI/FRII break ($\sim 5\cdot 10^{25}$ W
Hz$^{-1}$, Fanaroff \& Riley 1974\nocite{fr74}). In paper I it is
shown that this is also the case for the DDRGs with the exception of
B\,1834+620.

Whether the rate at which the black holes in AGN accrete and in the
process launch jets is limited by the availability of fuel or
otherwise is unclear. In any case it is unlikely that the efficiency
of the jet production mechanism will decrease if an additional gas
supply becomes available to the accretion disk if other parameters
like the mass and spin of the black hole do not change. If the spin of
the central black hole is responsible for the jet production
(e.g. Blandford \& Znajek 1977\nocite{bz77}) than we expect that the
new jets forming after disruption of the jet flow will have the same
power as the old jets.

\subsection{The evolution of FRII radio sources}
\label{sec:evo}

To investigate the radio properties of the DDRGs we will use the
dynamical model for FRII sources by Kaiser \& Alexander
(1997\nocite{ka96b}, hereafter KA) with the extension by Kaiser,
Dennett-Thorpe \& Alexander (1997, hereafter KDA)\nocite{kda97a} which
allows the calculation of the radio luminosity of these objects as a
function of their physical, linear size, $D$.

The observation that the optical identifications of the DDRGs are
 extended and the absence of broad lines in the optical spectra of the
 host galaxies (paper I) strongly suggest that the radio axes of the
 DDRGs lie close to the plane of the sky. We note that the sources
 3C445 and 3C 219 are broad-line objects and thus possibly oriented
 differently. But, since we do not take these two sources into account
 here, we will therefore assume that the projected linear size of the
 outer sources structures is indeed equal to their physical size. The
 inner source structures are very closely aligned with the outer
 structures. Although this may be a projection effect and the radio
 axes of the inner and outer structure may form a rather large angle,
 it is unlikely that all the DDRGs found until now conspire to produce
 the apparent very close alignment of the two parts of the source. We
 therefore assume in this paper that our viewing angle of the inner
 source structures is also very close to 90$^{\circ}$. In any case,
 all DDRGs are radio galaxies and orientation unification schemes
 suggest that the smallest viewing angle for a radio galaxy is
 45$^{\circ}$ (e.g. Barthel 1989\nocite{pb89}) which implies a maximum
 error in the physical linear sizes of $\sqrt{2}$.

The models of KA and KDA are based on the assumption of a constant
energy transport rate, $Q_o$, from the core of the radio galaxy via
the jets to the cocoon. The jets end in strong shocks which can be
identified with the radio hot spots and the jet material subsequently
inflates the radio cocoon. The expansion of the cocoon is supersonic
with respect to the surrounding material and therefore drives a bow
shock into this gas (Scheuer 1974\nocite{ps74}). The jets are confined
by the pressure in the cocoon. Falle (1991)\nocite{sf91} showed that
the expansion of the bow shock should be self-similar and the model
presented in KA predicts self-similar growth of the cocoon as
well. This is supported by observations (e.g. Leahy \& Williams
1984\nocite{lw84}).

The model of KA requires the density distribution external to the
radio cocoon to be modeled by a power law; $n_x = n_o ( r /
a_o)^{-\beta}$, where $n_o$ is the density at a distance, $r$, of
one core radius, $a_o$, from the centre of the radio galaxy. This
power law is a good approximation to a King (1972)\nocite{ik72}
profile with central density $n_o$ and $\beta _{King} = \beta / 3$
outside a few core radii. With this assumption the age of a radio
source of linear size $D$ is given by

\begin{equation}
t = \left( \frac{D}{2 c_1} \right) ^{\frac{5-\beta}{3}} \, \left(
\frac{m_p n_o a_o^{\beta}}{Q_o} \right) ^{\frac{1}{3}},
\label{age}
\end{equation}

\noindent where $m_p$ is the mass of a proton and $c_1$ is a
dimensionless constant (see KA). For the pressure within the cocoon,
$p_c$, KA find

\begin{equation}
p_c \propto t^{(-4-\beta) / (5- \beta)}.
\label{pre}
\end{equation}

KDA develop a more sophisticated model of the cocoon which follows the
evolution of the population of relativistic electrons responsible for
the radio emission via the synchrotron process in the different parts
of the cocoon under the influence of energy loss processes. These
include adiabatic expansion, synchrotron radiation and inverse Compton
scattering of the cosmic microwave background radiation.

The model of KDA for the radio luminosity as a function of linear size
or age of the radio source also depends on the aspect ratio of the
cocoon, $R$, the Lorentz factor of the bulk velocity of the jet
material, $\gamma _j$, and the index of the energy distribution of the
relativistic electrons at the time of their injection into the cocoon,
$p$. Available low frequency radio maps indicate that $R=3$, the
median value found by Leahy \& Williams 1984\nocite{lw84} in their
sample of FRII sources, is reasonable for all the sources discussed
here (see paper I). The aspect ratio of the cocoon is determined by
the ratio of the pressure in the hot spot region and that within the
cocoon, $p_h/p_c$ (see KA). Kaiser \& Alexander (1998)\nocite{ka98b}
show that the assumption of ram pressure confinement of the cocoon
perpendicular to the jet axis overpredicts the value of $p_h/p_c$ and
we use their empirical fitting formulae instead. From this for $R=3$
we find $p_h/p_c \sim 8$ if $\beta =1.5$ and $p_h/p_c \sim 21.4$ if
$\beta =0$.

We assume only mildly relativistic flow in the jet, $\gamma _j =2$,
and for this Heavens \& Drury (1988)\nocite{hd88} show that $p=2.14$
at the jet shock. We set the ratios of specific heats of the jet
material and of the IGM surrounding the source to $5/3$ while those of
the cocoon material and of the energy density of the completely
tangled magnetic field within the cocoon are set to $4/3$. We also
follow KDA in assuming that the power law initially describing the
energy distribution of the relativistic electrons in the cocoon
extends to thermal energies and that the jet consists entirely of pair
plasma. The inclusion of protons in the jets changes the absolute
values of the quantities calculated in the following sections but has
no influence on any of the conclusions of this paper.

\subsection{The evolution of the old cocoon}
\label{sec:old}

The absence of hot spots in some of the outer cocoons of the DDRGs
implies that in these cases the old jets are no longer
active. Therefore the model discussed in the previous section is not
directly applicable to DDRGs since it assumes a constant supply of
energy to the cocoon by the jets. For simplicity we assume that the
jet power, $Q_o$, drops instantaneously to zero once the jet
production mechanism is disrupted. The last jet material accelerated
by the AGN just before the interruption occurs takes a time $t_t$ to
reach the hot spots in the old cocoon. Only after the last jet
material has passed through the old hot spots the evolution of the old
cocoon will start to deviate from the prediction of the models of KA
and KDA. Because of the relativistic bulk speeds in extragalactic
jets, we set $t_t \sim D/2 \, c$, where $D$ is the total linear size
of the outer cocoons and $c$ is the speed of light.

We will assume that the evolution of the pressure of the material in
the old cocoons and therefore also that of the energy density of the
magnetic field in this region during the time we are interested in is
given by the power law derived by KA (Eq. \ref{pre}), even after
the last jet material has reached the old cocoon. The information that
the jets have ceased to supply the old cocoon with energy will travel
through the old cocoon at the local sound speed. After roughly one
 sound crossing time the entire cocoon will continue to grow but now
this expansion is adiabatic since there is no further energy input
into the old cocoon. In the following we will justify this assumption
by showing that the sound crossing time for each of the sources
discussed here by far exceeds the time elapsed since the last jet
material reached the old cocoon until the time at which the source is
observed. Because the energy supply to the old cocoons has stopped,
the pressure in the old cocoons will decrease faster than assumed here
and therefore this analysis represents formally only an upper limit.

The assumption about the pressure in the old cocoon also implies that
the evolution of the overall size of the cocoons during this time is
indistinguishable from that of cocoons which are still supplied with
energy by their jets. The total age of the source is thus given by
Eq. (\ref{age}).

Once the last jet material has passed through the jet shocks at the
end of the old jets, the hot spots in the old cocoon will start to
disperse and blend into the rest of the cocoon material at the local
sound speed, $c_s$. Using the model of KA we find that $c_s$ is
typically of order $0.5 c$ in the hot spot region. For an upper limit
of the hot spot radius of 10 kpc we find that the hot spot will
disappear roughly within $7 \cdot 10^4$ yr; a fraction of the time it
takes the material in the jets of giant radio sources to travel from
the core to the hot spots. We can therefore neglect this time and we
will assume that the hot spots of the old cocoon disperse
instantaneously once the last jet material has passed through them.

The relativistic electrons and possibly positrons responsible for the
synchrotron emission observed in the cocoons of FRII radio sources are
probably accelerated in the strong shocks at the end of the jets
(e.g. Hargrave \& Ryle 1974\nocite{hr74}). Clearly this acceleration
process stops once the last jet material reaches the old cocoons. The
relativistic particles that were accelerated until this time will
loose their energy because of the energy loss processes discussed by
KDA. This model allows one to calculate the radio emission of parts of
the cocoon identified by their injection time into the cocoon and then
add up these various contributions by integrating over all injection
times. In our case, the integration is stopped at the injection time
of the last jet material. With the assumptions made above this allows
us to determine, for a given jet power, $Q_o$, by what factor $\Delta$
the radio luminosity of the old cocoons has dropped during the time
interval $t_{d}$ from the moment the last jet material is injected
into the old cocoons until the time at which we observe the source. A
lower limit for the jet power, $Q_{o,min}$, is given by the case in
which no dimming has taken place and we observe the source at a time
when the last jet material has not yet reached the old hot
spots. Lower values for $Q_o$ are not possible since these would imply
a source luminosity below that observed even without any dimming.

For the assumptions given in Sect. \ref{sec:evo} results for the
minimum jet power, $Q_{o,min}$, and the corresponding maximum age of
the outer cocoon, $t_{o,max}$, for the external density profile of
poor groups are given in Tab. \ref{tab:data}. Note that for B\,1240+389
$Q_{o,min}$ is less than $10^{37}$ W which is roughly the dividing
line between low power FRI-type sources and the more powerful FRIIs
(Rawlings \& Saunders 1991\nocite{rs91}, KA). However, it is very
likely that the old cocoon of B\,1240+389 has dimmed and that the power
of its old jets therefore was greater than $10^{37}$ W (see Section
\ref{sec:new}).

\begin{table*}
\caption{\footnotesize Source parameters for the double-double radio
sources from observations and derived from the models of KA and
KDA. $D_o$ and $D_i$ are the linear sizes of the outer and inner
source structure respectively. $S_o$ and $S_i$ are the core-subtracted
flux density measurements at 1.4 GHz for the outer and inner source
structure respectively. $z$ is the cosmological redshift (linear
sizes, flux measurements and redshifts are taken from paper I). $t_t$ is
the light travel time from the core of the radio source to the tip of
the old cocoon, i.e. roughly the time it takes the last jet material
to reach the end of the old cocoon after it leaves the
AGN. $Q_{o,min}$ is the minimum jet power required for the outer
cocoon assuming that it is still supplied with energy by the old
jets. $t_{o,max}$ is the corresponding age of the outer
cocoon. $t_{d,max}$ is the maximum length of time during which the old
cocoon could have dimmed to be still consistent with the
observations.}
{\centering
\begin{tabular}{lccccccccc}
& $D_o$ / kpc & $D_i$ / kpc & $S_o$ / mJy & $S_i$ / mJy & $z$ & $t_t$ / Myr & $Q_{o,min}$ / W & $t_{o,max}$ / Myr & $t_{d,max}$ / Myr\\
\hline
B\,0925+420 & 2450 & 803 & 99 & 63.7 & 0.365 & 4.0 & $8.1 \cdot 10^{37}$ & 321 & 35.1\\
B\,1240+389 & 860 & 320 & 24.1 & 7.8 & 0.30 & 1.4 & $8.2 \cdot 10^{36}$ & 203 & 32.1\\
B\,1450+333 & 1680 & 180 & 426 & 33.5 & 0.249 & 2.7 & $7.3 \cdot 10^{37}$ & 214 & 34.5\\
B\,1834+620 & 1660 & 428 & 604 & 200 & 0.519 & 2.7 & $3.7 \cdot 10^{38}$ & 122 & 17.7\\
4C\,26.35 & 730 & 197 & 962 & 66.5 & 0.112 & 1.2 & $1.9 \cdot 10^{37}$ & 127 & 26.4\\
\hline
\end{tabular}}
\label{tab:data}
\end{table*}

\subsection{The evolution of the new cocoon}
\label{sec:new}

The inner source structure of all the DDRGs discussed here is fairly
symmetrical about the core of the respective source. The difference of
the lengths of the inner radio lobes on the two sides of a given
source (paper I) is comparable with typical values found for radio
galaxies without double-double structure (McCarthy, van Breugel \&
Kapahi 1991\nocite{mbk91}). This implies a density distribution in
their surroundings as smooth as that of sources embedded in the
unperturbed IGM. If the material in this region responsible for the
development of jet shocks in the inner source structures were clumpy,
we would expect the new jets to increase in luminosity if they
encounter a dense clump of gas; the propagation of the hot spots is
correspondingly slowed down. In all the DDRGs discussed here hot spots
are detected on both sides of the inner source structure at similar
distances from the core of the source and the ratio of the radio
luminosity of the two sides (paper I) is also within the limits found
for FRII radio galaxies without double-double morphology (McCarthy et
al. 1991\nocite{mbk91}). The ratio of the luminosities for B\,1450+333
($\sim 6$) is somewhat higher than in the other sources. However note,
that in this source the lengths of the two sides of the inner source
structure are almost identical. The armlength ratio is 1.06. We
therefore consider it likely that the environment of the inner source
structures of DDRGs can also be modeled by a power law density
distribution similar to what is assumed for `normal' FRIIs (see KA).

For all five aligned DDRGs the direction of the new jets is within
7$^{\circ}$ of the old jet axis. This implies that in each object the
inner structure is expanding in a region occupied by the material of
the old cocoons. The new jets all end in hot spots and the gas
surrounding the inner source structures must therefore be dense enough
to cause the formation of strong jet shocks.

Clarke \& Burns (1991)\nocite{cb91} present numerical simulations of a
restarting jet. They find that their simulated jet develops a shock
within the region of the old cocoon indicating that the gas within the
old cocoon is dense enough to prevent the new jet becoming
`quasi-ballistic', i.e. with only a negligible pressure discontinuity
(shock) at its end. The shock they observe in their simulations may be
strong enough to cause the inner source structure of the aligned DDRGs
discussed here.

Although Clarke \& Burns find that mechanical instabilities along the
cocoon boundary begin to grow after the jet is `switched' off, the gas
causing the formation of a jet shock is mainly that transported by the
old jet during its activity. They find the resulting density contrast
of the gas in the cocoon and the uniform density the old jet expanded
into to be roughly $1/40$. The model of FRII sources used here assumes
that no mixing of material across the contact discontinuity
delineating the cocoon of these objects takes place. For this case the
only material within the old cocoons of the DDRGs is the jet material
which has passed through the jet shock. The density in the old cocoon
in the case of a uniform distribution is given by the model as

\begin{equation}
n_c = \frac{Q_o \left( t_o -t_{d}\right)}{\left( \gamma _j
-1 \right) m_p c^2 V_c},
\label{rhoc}
\end{equation}

\noindent where $t_o$ is the total age of the source given by Eq.
(\ref{age}), $t_d$ is the `dimming time', i.e. the time elapsed
between the arrival of the last jet material at the old cocoon and the
time of observation, and $V_c$ is the volume of the old cocoon. In
order to be able to compare the density given by Eq. (\ref{rhoc})
with the density of the unperturbed IGM we have assumed that all the
particles within the old cocoon are protons. This is contrary to our
assumption that the jets consist of electrons and positrons
only. However, since all model predictions for the evolution of the
inner source structure depend only on the mass density and not on the
particle density in the old cocoon, we can use Eq. (\ref{rhoc})
for purposes of comparison in its given form. If we assume that the
old cocoon expanded in a uniform density environment ($n_o \sim
10^{-2}$ cm$^{-3}$, $\beta =0$) as in the numerical simulations of
Clarke \& Burns (1991)\nocite{cb91}, than we find from Eq.
(\ref{rhoc}) that our model predicts density contrasts of the gas in
the cocoon and the ambient medium comparable to those found in the
simulations for short lengths of the old cocoon (of order 1 kpc). This
is consistent with the numerical simulations since they only extend to
short life times equivalent to short lengths of the jet. 

Clarke \& Burns (1991)\nocite{cb91} also point out that the formation
of relatively strong shocks at the end of the inner jets should be
accompanied by the formation of a bow shock within the old cocoon
material of similar strength in front of the hot spots. Since the old
cocoon volume is filled with magnetic fields this bow shock could be
visible due to synchrotron emission if electrons are accelerated to
relativistic velocities by the bow shock. In none of the sources
discussed here a distinct bow shock around the inner source structure
has been observed. This may indicate that the acceleration of
particles to relativistic velocities at this bow shock is not
efficient. Note here that the strength of the bow shock around the
inner source is decreasing away from the hot spots. This implies that
although the shock ending the inner jet is strong enough to accelerate
particles `lighting up' the cocoon of the inner sources the associated
bow shock may be too weak along most of its length to produce enough
relativistic particles for it to be detectable. However, the
non-detection of any emission from this bow shock is somewhat
puzzling.

In the case of short life times of the old jets before the jet flow is
interrupted and of uniform density distributions of the external gas
we showed above that the gas in the region of the old cocoon may
indeed be dense enough to force the formation of significant shocks at
the end of the new jets. However, for the profile of the density in
the environment used here and for the considerably larger life times
of the old jets derived in the previous section, Eq. (\ref{rhoc})
predicts much smaller density contrasts. We will show in the following
that the density in the region of the old cocoon created by the
material transported by the old jets alone is insufficient for the
formation of strong shocks at the end of the new jets in DDRGs.

\begin{figure*}
\centerline{\epsfig{file=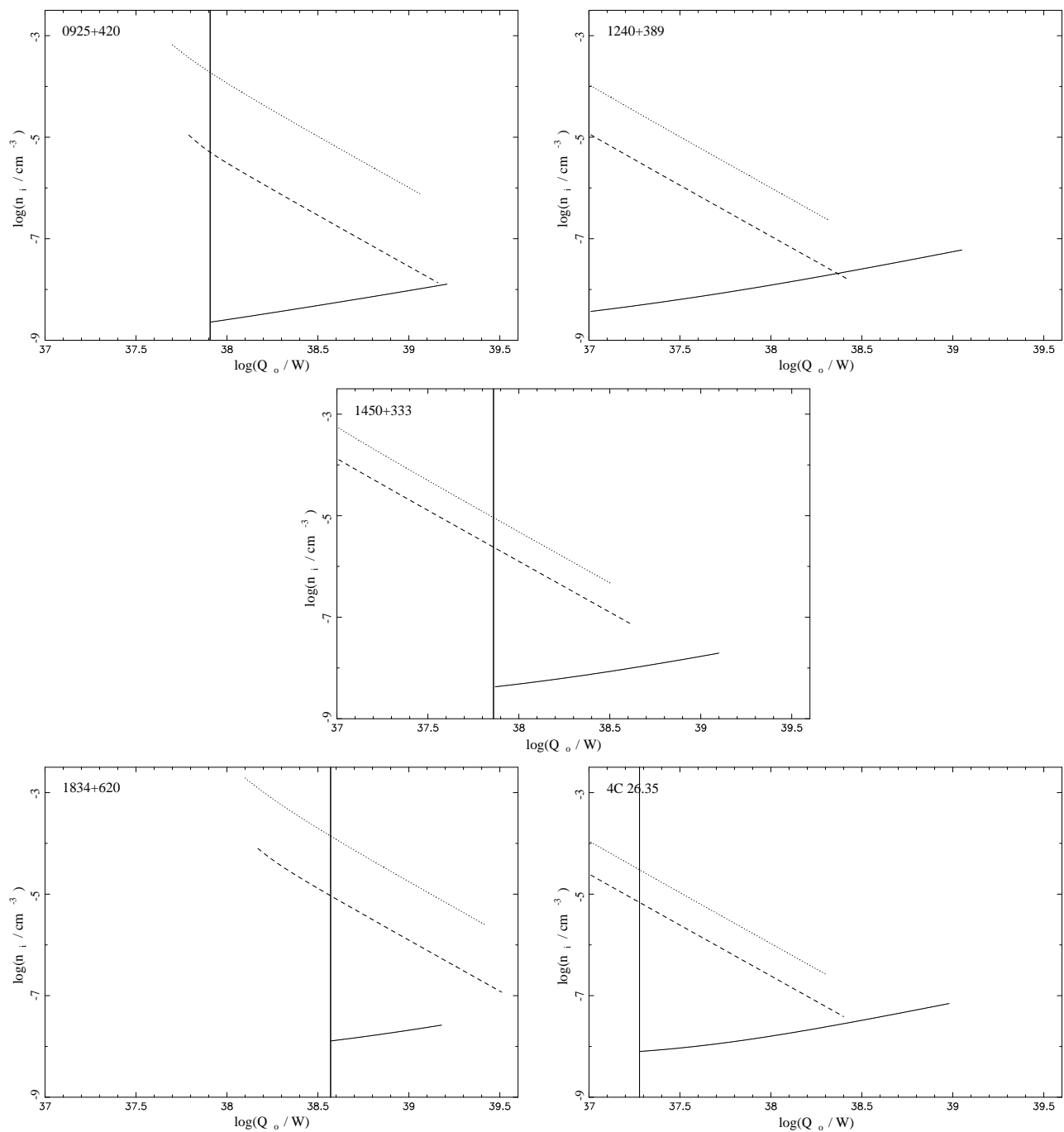, width=17cm}}
\caption{\footnotesize Densities within the region of the old cocoon
of the DDRGs. Vertical solid line: $Q_{o,min}$ of the jets which
inflated the old cocoon. Solid line: The uniform density produced in
the old cocoon by the material of the old jet alone (Eq.
\ref{rhoc}) as a function of the power of the old jets. Dashed line:
The uniform density required to produce the observed properties of the
inner source as a function of the power of the inner jets. Dotted
line: The central density, $n_{oi}$ required within the old cocoon
region if $\beta _i =1.5$ and $a_o=10$ kpc.}
\label{fig:rhoi}
\end{figure*}

Numerical simulations predict the presence of backflow in the cocoon
(e.g. Norman et al. 1982)\nocite{nsws82} which implies that most of
the material within the cocoon will `pile up' towards the core of the
source. This suggests that the new jets are propagating along a
negative density gradient, the exact shape of which is difficult to
determine. However, an upper limit for the density in front of the hot
spots of the new source is given by the assumption that all the jet
material which has passed through the old jets during their life time
is uniformly distributed over the volume of the old cocoon. If, within
the confines of the inner source, the same amount of material is
arranged in such a way that it forms a monotonically decreasing
density profile, the density in front of the hot spots of the inner
source will always be less than in this limiting case. For calculation
of the density in the region of the old cocoon in this limiting case
we can therefore use Eq. (\ref{rhoc}). The solid lines in Fig.
\ref{fig:rhoi} show the results as a function of the power of the old
jets. We did not consider jet powers above a value for which the radio
luminosity at 178 MHz of the respective source would reach $2\cdot
10^{27}$ W Hz$^{-1}$ sr$^{-1}$, if no dimming of the emission had
taken place. This is the luminosity of the most luminous source with a
linear size greater than 700 kpc, 3C 292, in the sample of Laing,
Riley \& Longair (1983)\nocite{lrl83} which includes the most luminous
radio sources in the observable universe at any redshift.

\begin{figure*}
\centerline{\epsfig{file=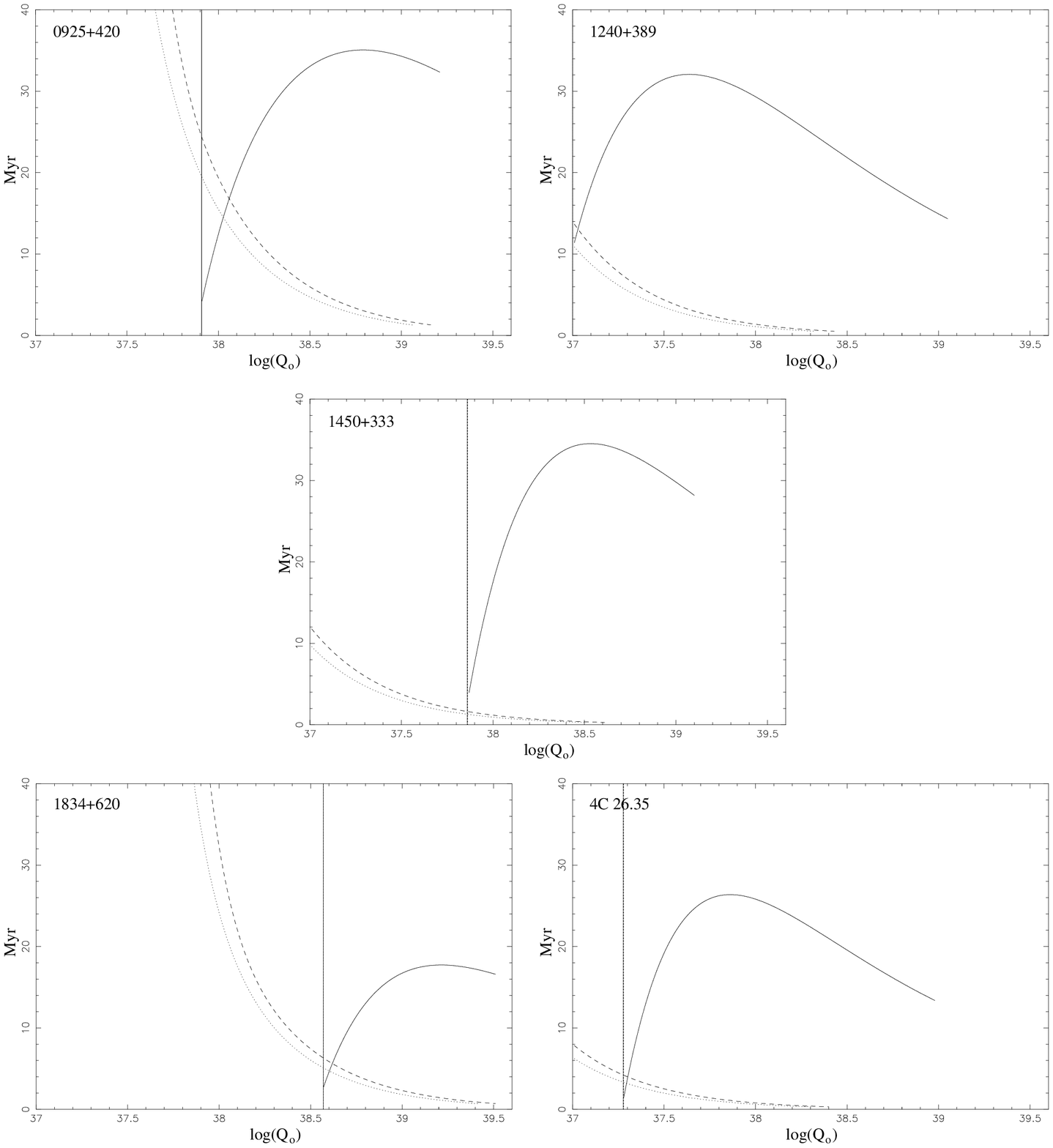, height=18cm}}
\caption{\footnotesize Time scales in the DDRGs. Solid vertical line:
The minimum jet power derived from the model of KDA (see text and
Tab. \ref{tab:data}). Solid line: The sum of $t_t$ and $t_{d}$ as a
function of the power of the old jets. $t_t$ is given by the
intersection of this curve with the solid vertical line. Dashed line:
The age of the inner source structure as a function of the power of
the inner jets for the assumption of a uniform density in the
environment. Dotted line: The age of the inner source structure for
$\beta _i =1.5$.}
\label{fig:toff}
\end {figure*}

Analogous to the analysis of the old cocoon, we can use the model of
KDA to determine the density of a uniform environment, $n_{oi}$,
required to produce the observed linear sizes and radio fluxes of the
inner sources for a given jet power. We use the same source parameters
as for the old cocoon except for the exponent of the external density
which in a uniform environment $\beta _i =0$ (note that the exponent
of the power law density distribution surrounding the inner source
structure, $\beta _i$, is in general not equal to $\beta$, the
equivalent exponent of the material the outer cocoon is embedded
in). The dashed lines in Fig. \ref{fig:rhoi} show $n_{oi}$ as a
function of the power of the inner jets. The cut-off at high jet
powers for these lines in the figure is given by the limit that the
inner source can not expand faster than the speed of light. It is now
also possible to calculate the age of the inner source, $t_i$, as a
function of the jet power of the inner source. This is shown as the
dashed lines in Fig. \ref{fig:toff}.

The time during which the AGN was quiescent, $t_{off}$, is given by 

\begin{equation}
t_{off} = t_t + t_{d} -t_i,
\end{equation}

\noindent where $t_t$ is the time it takes the last jet material to
reach the end of the old cocoon and $t_d$ is the time during which the
old cocoon must have dimmed for a given power of the old jets in order
to explain the currently observed properties of the outer radio
lobes. The solid lines in Fig. \ref{fig:toff} represent the sum of
$t_t$ and $t_d$ as a function of the power of the old jet. The turn
over in these curves is caused by a change in the dominant energy loss
process for the relativistic electrons in the old cocoon. For lower
jet powers this is the inverse Compton scattering of the CMBR while at
higher $Q_o$ losses due to synchrotron radiation become more
important. The maximum dimming time, $t_{d,max}$, for each source is
given by the peak of the solid curves in Fig. \ref{fig:toff} minus
$t_t$. They are summarized in Tab. \ref{tab:data}. 

$t_{off}$ must always be greater or equal to zero and this implies
that $t_i \le t_t + t_{d,max}$. This condition defines a minimum jet
power for the new jets which is shown as the cut-off of the dashed
lines in Fig. \ref{fig:rhoi} at low jet powers for two of the
DDRGs. For the other three sources this cut-off is below $10^{37}$ W
which is roughly the jet power where the transition from FRII to
FRI-type sources occurs (Rawlings \& Saunders 1991\nocite{rs91}, KA).

From Fig. \ref{fig:rhoi} it is clear that in the cases of B\,0925+420,
B\,1240+389 and 4C\,26.35 the properties of the inner source structure
could be caused alone by the presence of the material transported by
the old jet during its life time. B\,1240+389 and 4C\,26.35 are also the
sources with the smallest outer structures in our sample for which
Eq. (\ref{rhoc}) predicts the highest densities in the old
cocoon. Note however, that for all three objects this requires the
somewhat favorable assumption that this material is distributed
uniformly. It also implies that even if the power of the new jets in
B\,1240+389 and 4C\,26.35 is lower by a factor of 5 to 10 than that of the
old jets the current expansion velocity of the inner structure is very
close to the speed of light, which is unlikely. We therefore conclude
that the old cocoons of the DDRGs can not be filled solely with the
old jet material but that they must be `dirty' in the sense that some
other material must have penetrated the cocoon boundary from the
outside.

If the power of the new jets in B\,1450+333 is equal to that of the old
jets, then the observations of this source are consistent with the old
cocoon still being supplied with energy. This would imply the
existence of hot spots in the old cocoon of this source. The
observational evidence is unclear, but a rather diffuse hot spot may
have been detected in the southern radio lobe (paper I). In this
scenario we find $t_{off} \sim 10^6$ years for B\,1450+333. Assuming that
the power of the inner jets, $Q_o$, is equal to that of the old jets
as suggested in Sect. \ref{sec:disr} and that $t_{off} \sim 1$ Myr
for all sources, we can calculate $Q_o$ required for each source by
the observed properties of the inner and outer source structure. Tab.
\ref{tab:res} lists the results for the assumption of a uniform
density in the region of the old cocoon along with the required
densities and resulting source ages. The assumption of $t_{off}=1$ Myr
gives plausible results for $Q_o$ but other values for $t_{off}$ may
be reasonable as well. B\,1834+620 shows a radio hot spot at one end of
its old cocoon but not at the other (paper III). This implies that the
radio emission of the old cocoon in this source has not dimmed
significantly and the small value for $t_d$ listed in Tab.
\ref{tab:res} for the assumption of $t_{off} =10^6$ years for this
source is within the limitations of the model consistent with this.

The sound crossing time for the old cocoon, $t_{sc}$, is also given in
Tab. \ref{tab:res}. For all five sources the time during which the
radio emission of the cocoon has dimmed is of order a few percent of
the sound crossing time. Our assumption that most of the old cocoon
continues to dynamically evolve during $t_d$ as if the jets were still
supplying it with energy is therefore justified. 

\begin{table*}
\caption{\footnotesize Derived source parameters for the assumption of
constant jet power, uniform density within the region of the old
cocoon and $t_{off} \sim 10^6$ years. $Q_o$ is the jet power of the
old and the new jets. $n_{oi}$ is the density within the old cocoon
region. $f_{cl}'$ is the initial filling factor of line emitting
clouds within the old cocoon necessary to produce this density if all
of the cloud material is spread over the volume of the cocoon. $t_i$
is the age of the inner source structure. $t_o$ is the age of the
outer source structure. $t_d$ is the time during which the old cocoon
has dimmed. $\Delta$ is the factor by which the radio emission of the
old cocoon has dimmed since the last jet material reached the old hot
spots. $t_{sc}$ is the sound crossing time of the old cocoon for the
density $n_{oi}$.}  
{\centering
\begin{tabular}{lcccccccc}
& $Q_o$ / W & $n_{oi}$ / cm$^{-3}$ & $f_{cl}'$ & $t_i$ / Myr & $t_o$ / Myr & $t_d$ / Myr & $\Delta$ & $t_{d}/t_{sc}$\\
\hline
B\,0925+420 & $1.2 \cdot 10^{38}$ & $1.3 \cdot 10^{-6}$ & $1.3 \cdot 10^{-8}$ & 16.0 & 281 & 13.0 & 1.8 & 0.033\\
B\,1240+389 & $1.1 \cdot 10^{37}$ & $5.8 \cdot 10^{-6}$ & $6.0 \cdot 10^{-8}$ & 12.8 & 184 & 13.3 & 1.5 & 0.054\\
B\,1450+333 & $7.2 \cdot 10^{37}$ & $1.4 \cdot 10^{-6}$ & $1.4 \cdot 10^{-8}$ & 1.6 & 215 & 0.0 & 1.0 & 0.0\\
B\,1834+620 & $4.3 \cdot 10^{38}$ & $4.1 \cdot 10^{-6}$ & $4.0 \cdot 10^{-8}$ & 5.4 & 117 & 3.7 & 1.2 & 0.017\\
4C\,26.35 & $2.0 \cdot 10^{37}$ & $3.5 \cdot 10^{-6}$ & $3.5 \cdot 10^{-8}$ & 3.9 & 125 & 4.2 & 1.1 & 0.037\\
\hline
\end{tabular}}
\label{tab:res}
\end{table*}

\section{`Contaminating' the old cocoon}
\label{sec:cont}

Several possibilities for the contamination of the cocoon with
material from the outside exist. In this section we consider
entrainment of material into the old cocoon across the contact
discontinuity by Kelvin-Helmholtz and/or Rayleigh-Taylor
instabilities, the replacement of the old cocoon by the surrounding
IGM by buoyancy and the disruption and dispersion by the bow shock of
warm, dense clouds embedded in the IGM.

\subsection{Entrainment across the contact discontinuity}

Numerical simulations indicate that the contact discontinuity
delineating the cocoon is stable against hydrodynamic instabilities,
if the backflow within the cocoon is supersonic with respect to the
ambient medium (Norman et al. 1982)\nocite{nsws82}. In the simulations
the backflow is found to be initially supersonic whenever the bulk
velocities in the jets inflating the cocoon are highly supersonic with
respect to the speed of sound in the unperturbed IGM. For the five
DDRGs and the assumed isothermal density profile of poor groups of
galaxies we find that even if the power of the old jets is close to
their lower limit, $Q_{o,min}$, their Mach numbers just before they
stop supplying energy to the old cocoon are equal or greater than
about 5.

In FRII sources which do not show large off-axis gas flow within their
cocoon the backflow of gas must eventually be decelerated to
velocities below the external sound speed (Norman et
al. 1982\nocite{nsws82}). This may lead to the development of fluid
instabilities along the contact discontinuity of the inner
cocoon. This is also seen in numerical simulations. However, most
simulations are confined to a two dimensional treatment of the problem
and include reflective boundary conditions at the limits of the
computational grid where we would expect the instabilities to be
strongest. The predictions of numerical simulations in this flow
region should therefore be treated with caution. A simple analytical
estimate of the time scale for the growth of Kelvin-Helmholtz and
Rayleigh-Taylor instabilities of length scale $l$ at the boundary of
two fluids with large density contrast $\chi$ and relative velocity
$v_{rel}$ is given by $t_{KH} \sim \sqrt{\chi} \, l / v_{rel}$ and
$t_{RT} \sim \sqrt{l/g}$, where $g$ is the acceleration of one fluid
with respect to the other (Chandrasekhar 1961\nocite{sc61}). To
replace large parts of the cocoon of FRII sources large-scale
instabilities would be most efficient. However, for $l$ comparable to
the cocoon size the instability growth is slow and the mixing of
material from the ambient IGM with the gas in the cocoon on small
scales, say $l\sim 1$ kpc, may be sufficient if it proceeds fast
enough. We have mentioned already that the relative velocity for the
two gas streams for fluid instabilities to grow must be of the order
of or smaller than the sound speed in the unshocked IGM, which for
$T_x \sim 10^7$ K is about 370 km s$^{-1}$. From the previous section
we note that $\chi \sim 10^4$ for the outer cocoon at a few 100 kpc
from the centre of the assumed density distribution of the unshocked
IGM. From this we find that $t_{KH} \sim 2 \cdot 10^8$ years which is
longer than the life time of most of the outer cocoons in the sources
discussed here. Kaiser \& Alexander (1999)\nocite{ka98b} show that
there should be a backflow not only within the cocoon but also in the
shocked IGM in between bow shock and contact discontinuity. Both flows
show similar velocities and decelerate on similar length scales. This
implies that $g$, the acceleration of the flow within the cocoon with
respect to that of the shocked IGM, is small and $t_{RT} \sim
t_{KT}$. Entrainment of dense material from the IGM into the old
cocoon across the contact discontinuity even in the regions of the
cocoon where the backflow velocity is not supersonic with respect to
the unshocked IGM should therefore be rather inefficient. This makes
it unlikely that the additional material in the old cocoons has been
entrained.

\subsection{Replacement of the old cocoon by the IGM}
\label{sec:rep}

During most of the life time of FRII sources the cocoon will be
overpressured with respect to the surrounding IGM. The expansion of
the cocoon will be supersonic and therefore it will drive a strong bow
shock into the IGM. However, the pressure within the cocoon decreases
with time and will eventually become comparable to the pressure of the
ambient medium and the bow shock will vanish. If the IGM in the
vicinity of the cocoon is isothermal with a density distribution close
to a King (1972)\nocite{ik72} profile, this will first occur close to
the centre of the radio source. Once the cocoon is no longer protected
by its bow shock over its entire length, buoyancy will set in and the
denser IGM will push the lighter material filling the cocoon outwards
thereby starting to replace it in the inner regions of the
cocoon. Assuming the minimum jet power for each source given in Tab.
\ref{tab:data}, the model of KA predicts that for a temperature of
$10^7$ K for the IGM the pressure in the old cocoons of all of the
DDRGs discussed here, except in that of B\,1834+620, should be slightly
lower than about half the value of the ambient pressure close to the
core of the respective source. In the case of B\,1834+620 the old cocoon
should still be overpressured with respect to the IGM over its entire
length. The low value for the pressure in the old cocoon in the other
sources implies that their cocoons may be in direct contact with the
unshocked IGM close to the core. However, this estimate depends
strongly on the assumed temperature of the IGM and also on whether
this material really is isothermal. If the power of the old jets is
higher than their minimum value used here, then their old cocoons will
still be overpressured and they will still be surrounded by a bow
shock over their entire lengths.

The replacement of the cocoon material by the IGM due to buoyancy
proceeds at the sound speed of the IGM and is therefore a slow
process. For the temperature of the IGM assumed here, the relevant
sound speed is of order 370 km s$^{-1}$. At this speed we find that
the replacement of a cylindrical volume of 100 kpc length and the
width of the old cocoon by the IGM takes of order a 100 Myrs. This
almost corresponds to the maximum ages of the old cocoons (see Tab.
\ref{tab:data}). This implies that there is enough time for the IGM to
replace a considerable fraction of the volume of the old cocoon only
if the replacement started early in the evolution of the radio
source. Of course, a higher temperature of the IGM causing a higher
sound speed in this material would shorten the relevant time
scales. Although in principle this mechanism may therefore be fast
enough we present further arguments against the replacement of large
fractions of the cocoon material by the IGM in the following.

It is not straightforward to determine the density distribution in the
region of the old cocoon resulting from such a buoyant replacement of
the old cocoon material. We can, however, make two very crude
approximations. If the IGM has had enough time to replace the old
cocoon material and settle into an equilibrium configuration, its
density distribution may closely resemble that of the unperturbed
IGM. In this case the inner sources in the aligned DDRGs are embedded
in essentially the same environment as the old cocoons with $n_{oi}
=10^{-2}$ cm$^{-3}$, $a_o = 10$ kpc and $\beta _i =1.5$. The dotted
lines in Fig. \ref{fig:rhoi} show the central density $\rho _{oi}$
required to explain the properties of the inner source structures for
the case of $\beta _i=1.5$ assuming that $a_o=10$ kpc. The density
required to explain the observations is well below the value assumed
for this approximation. If the entire material of the IGM contained in
a sphere of radius 400 kpc, approximately the linear size of one half
of the cocoon of the inner source structure of B\,0925+420, centered on
the core of the radio source is distributed uniformly over the same
volume by the replacement process, the density in this region is found
to be $\sim 10^{-4}$ cm$^{-3}$. Again this is much higher then the
density required for the inner source structures in the case of a
uniform density profile within the region of the old cocoon (dashed
lines in Fig. \ref{fig:rhoi}). Both approximations indicate that the
replacement of the old cocoon material with the denser IGM can not
explain the properties of the inner source structures.

If the replacement of the cocoon material takes place in combination
with mixing of this gas with the IGM this may result in the low
densities required for environments of the inner source
structures. However, as we have shown in the previous section, the
time scales for the fluid instabilities that could be responsible for
mixing of the two fluids are prohibitively long. 

Kaiser \& Alexander (1998)\nocite{ka98b} discuss the `pinching' of the
cocoon at its centre by a steep pressure gradient within the
cocoon. This gradient is caused by steep gradients in the density
distribution of the surrounding IGM. This effect could also be
responsible for the replacement of the old cocoon material but
analogous to the replacement due to buoyancy it is not clear how the
replacing IGM is diluted sufficiently to give rise to the observed
properties of the inner sources.

Taking together all the arguments above it seems unlikely that the
environments of the inner sources of the DDRGs are created by the
replacement of the old cocoon material by the denser IGM.

\subsection{Dispersion of warm clouds}
\label{sec:disp}

As we have seen in Sect. \ref{sec:enviro}, optical observations of
FRII radio galaxies suggest the existence of warm, dense clouds
embedded in the hot gas of the IGM in the vicinity of the host
galaxy. They can potentially provide the additional material needed to
explain the observed properties of the inner source structures in the
aligned DDRGs. This is only possible if they can pass through the
contact discontinuity into the old cocoon and if most of their
material is subsequently dispersed over the volume of the old cocoon.

When the bow shock of an FRII radio source encounters one of these
warm clouds embedded in the hot IGM, it drives a shock into it. The
cloud is small and will therefore quickly re-establish pressure
equilibrium with its surroundings. This implies that the Mach number
of the shock within the cloud can be set equal to the Mach number of
the bow shock within the hot IGM, $M_b$ (McKee \& Cowie
1975\nocite{mc75}). Note that because of its shape $M_b$ is not
uniform along the length of the bow shock. In the following $M_b$
therefore refers to an `average' Mach number of a given bow
shock. Because of its large density the cloud is not efficiently
accelerated by the passage of the shock and is therefore quickly
overtaken by the contact discontinuity and passes through this surface
into the cocoon.

It has been suggested that after the cloud has passed through the bow
shock it will collapse and possibly form stars if radiative cooling is
efficient (Rees 1989\nocite{mr89}, Begelman \& Cioffi
1989\nocite{bc89}, Foster \& Boss 1996\nocite{fb96}). Because of the
low temperature of the cloud material the bow shock of FRII sources
will be radiative and the compression of the cloud is an almost
isothermal process. For strong shock conditions this implies $n_{cl}'
= M_b^2 n_{cl}$, where $n_{cl}$ is the pre-shock density of the cloud
(Rees 1989\nocite{mr89}). We will use dashed variables for quantities
describing the cloud after the shock has passed through it. The
temperature of the warm cloud will initially increase by a factor
$\sim 30$ for $M_b =10$. The cloud material then cools radiatively
back to a temperature of $10^4$ K within $10^3 \rightarrow 10^4$ years
(Sutherland \& Dopita 1993\nocite{sd93}). During this phase the
optical line emission of the clouds will be enhanced because of the
passage of the bow shock. Once the temperature within the cloud
reaches $10^4$ K, cooling via radiation becomes inefficient and the
subsequent evolution of the cloud will be adiabatic. In this regime
the numerical simulations predict the shredding of the cloud due to
growing Kelvin-Helmholtz and Rayleigh-Taylor instabilities on its
surface within a few `cloud crushing times' (Klein, McKee \& Colella
1994\nocite{kmc94})

\begin{equation}
t_{cc} = \frac{\sqrt{\chi'} r_{cl}'}{v_b},
\label{crush}
\end{equation}

\noindent where $\chi'$ is the ratio of the density of the warm cloud
and that of its surroundings, $r_{cl}'$ is the radius of the cloud and
$v_b$ is the velocity of the bow shock with respect to the unshocked
IGM. Here we have made use of the fact that the warm cloud is not very
effectively accelerated by the passage of the bow shock and that $v_b$
is therefore similar to $v_{rel}$, the relative velocity of the
shocked hot IGM with respect to the cloud.

For this and in the following we assume that the density of the cloud
is uniform over its volume. This is a simplification and the cloud
will be denser close to its centre than further out. In this case the
core of the cloud may collapse under the influence of the radiative
shock within the cloud and form stars. Such shock induced star
formation may play an important role in the explanation of the
alignment effect in FRII sources at $z > 0.6$ (see Sect.
\ref{sec:opt}). However, most of the gas in the cloud will become
adiabatic before it can collapse to form stars and this is the
material which will subsequently be spread out over the volume of the
old cocoon. Super novae and stellar winds of the stars formed in the
core of the cloud will contribute to the dispersion of the outer cloud
regions as well. The exact details of the evolution of the warm cloud
can probably only be determined with the help of numerical simulations
taking into account radiative cooling and feedback from star
formation.

Immediately after passing through the bow shock the warm cloud is
surrounded by the shocked gas of the hot IGM. For the hot IGM the bow
shock is adiabatic and for strong shock conditions $\chi'$ is
therefore equal to $\sim M_b^2/4 (T_x/T_{cl})$, where $M_b$ is the
local Mach number of the bow shock. This implies the shredding of the
cloud within a few $t_{cc} \sim M_b \, \sqrt{T_x/T_{cl}}\, r_{cl}'/(2
c_x)$. The acceleration of the cloud by the shocked IGM outwards, away
from the contact discontinuity, proceeds on at least similar if not
longer time scales (Klein et al. 1994\nocite{kmc94}). Consider now a
`typical' FRII source with a total linear size of $D=400$ kpc embedded
in a density distribution appropriate for a poor group. According to
Eq. (\ref{age}) this source has an age of roughly $2\cdot 10^7$
years if the aspect ratio of its cocoon, $R$, is equal to 3. From this
we find the advance speed of the hot spots for this source $v_b \sim
0.03$ c. If we assume that the shape of the cocoon is cylindrical the
expansion speed perpendicular to the jet axis is $5\cdot 10^{-2}$ c
thereby implying a Mach number of the bow shock in this direction of
3.8. Assuming $r_{cl}' = 10$ pc (e.g. Osterbrock 1989\nocite{do89}) we
find that for this cloud in these conditions $t_{cc} \sim 1.6 \cdot
10^6$ years. From Kaiser \& Alexander (1999)\nocite{ka98b} we note
that the stand-off distance between the bow shock and the contact
discontinuity is roughly $5\cdot10^{-3} D$. If the cloud is not
accelerated at all by the passage of the bow shock, it will take the
contact discontinuity roughly $10^6$ years to overtake the cloud. The
cloud is therefore just able to reach the safety of the cocoon before
it is dispersed within the layer of shocked IGM. Since the shape of
the cocoons of FRII sources is not cylindrical, the conditions for
clouds closer to the hot spot should be more favorable than estimated
here because the Mach number of the bow shock will always be higher
than in the above scenario and also the different direction of
expansion of the cocoon causes the contact discontinuity to overtake
the cloud earlier. We therefore conclude that `average' clouds are not
disrupted or accelerated efficiently by the bow shock and the shocked
IGM.

The pressure within the cocoon is the same as that in the shocked
layer of gas between bow shock and contact discontinuity but the
density is much lower. For an estimate we use Eq. (\ref{rhoc})
with the same parameters as above. Again we assume that the warm
clouds supplying the material for the environment of the inner source
structures are not displaced by the passage of the bow
shock. Therefore to determine the disruption time scale of these
clouds we have to consider the properties of the bow shock around the
outer source structure at the time when it had a linear size of
typically 400 kpc. We find $n_c \sim 1 \cdot 10^{-7}$ cm$^{-3}$ which
implies $t_{cc} \sim 5 \cdot 10^7$ years. A warm cloud is therefore
stable for a long time within the cocoon during which its outer layers
may be ionized by the UV emission of the central AGN and/or the
emission of the shocked hot phase of the IGM (see Sect.
\ref{sec:opt}). This material may then produce the observed line
emission.

Roughly a time $t_{cc}$ after an ensemble of clouds have passed
through the bow shock they are finally broken up into smaller
fragments and their material is spread out over large fractions of the
cocoon volume. It is not straightforward to estimate the size of the
volume over which the cloud material is dispersed because this is
determined by the mixing of the cloud material with the gas already
present in the cocoon. While the break-up of the warm clouds is mainly
driven by the fluid instabilities on large scales, mixing proceeds on
much smaller scales which are difficult to resolve in numerical
simulations (Klein et al. 1994\nocite{kmc94}). As was mentioned above,
instabilities grow faster on smaller scales and so significant
fractions of the gas mass of the clouds may be already mixed with the
other cocoon material by the time the clouds finally break up on
larger scales after a time $t_{cc}$. A rough estimate for the volume
over which the cloud material is spread may be obtained by calculating
the volume `swept-out' by a given cloud during one cloud crushing
time. Assuming that the cloud expands adiabatically in pressure
equilibrium with its surroundings in the cocoon we find from Eq.
(\ref{pre}) that the cloud cross sectional area, $\sigma _{cl}$,
increases proportional to $t^{2(4+\beta)/3 \Gamma _{cl}
(5-\beta)}$. If we again assume that the velocity of the cloud
relative to its surroundings is roughly equal to the advance speed of
the hot spot we find for the volume swept-out by the cloud $\int
\sigma _{cl} v_b \, dt$. If the cloud material is distributed over
this swept-out volume this corresponds to an increase in the cloud
volume of a factor $\sim 3\cdot 10^5$ for the cloud and source
parameters introduced above. Of course this estimate is very
simplistic and neglects many of the complicated processes
involved. However, we note that since the cloud disruption is a
continuous process the density of the gas in the cocoon responsible
for the disruption of newly incoming clouds is already enhanced by the
shredding of clouds which were earlier overtaken by the expansion of
the cocoon. This means that the time scales for the cloud dispersal
and mixing of cloud material with the cocoon gas may be shorter than
estimated above. From this we see that the volume filling factor of
the former cloud material may increase substantially when compared to
that of the intact clouds. In the following we will make the
approximation that the material of all clouds is spread out uniformly
over the entire volume of the cocoon, i.e. the filling factor of the
former cloud material is unity after the disruption of the clouds. The
resulting density within the cocoon is approximately $f_{cl}'
n_{cl}'$. This then provides the environment for the inner source
structures of the aligned DDRGs.

In Sect. \ref{sec:new} we have calculated the uniform density within
the region of the old cocoon required to explain the properties of the
inner source structure for the assumption that $t_{off} \sim 10^6$
years and that the jet power of the old and new jets is the same. If
this density profile is provided by the shredding of line-emitting
clouds we can estimate their initial volume filling factor, $f_{cl}'$,
within the cocoon before they are dispersed. The results which are
given in Tab. \ref{tab:res} agree very well with the filling factors
derived from observations of the line emission (Heckman et
al. 1991\nocite{hlmb91}). Note that because we assume a filling factor
of unity of the former cloud material the calculated values of
$f_{cl}'$ are only lower limits since some of the clouds will not be
dispersed completely. Of course this is also true if the cloud cores
collapse and form stars.

The distribution of warm, line-emitting clouds within the hot IGM must
be quite smooth to explain the symmetry of the inner source
structures. In some FRII sources observations indicate a rather clumpy
distribution of these clouds (e.g. Johnson, Leahy \& Garrington
1995\nocite{jlg95}). A concentration of a large number of warm, dense
clouds in the IGM in the path of one of the jet may even become
dynamically important for the evolution of this jet (McCarthy et
al. 1991\nocite{mbk91}). The volume filling factor of the clouds must
be locally increased in such regions so that $f_{cl}' \sim
\sqrt{T_x/T_{cl}} \ge 0.03$. However, the distribution of clouds in
these sources is observed at a time when they are still intact within
the cocoon. The time available for the cloud to disrupt and the cloud
material to smooth out over the volume of the old cocoon in DDRGs is
probably sufficient to provide a more homogeneous environment for the
inner source structures than suggested by the observations of the
intact clouds.

\section{Implications of cloud-shredding for aligned DDRGs and other large
FRII sources}
\label{sec:impl}

The dispersion of warm, dense clouds by the bow shock of the outer
cocoon of aligned DDRGs can explain the formation of the observed
peculiar structures of these objects. There are some further
implications of this model which we will discuss in this section.

\subsection{Internal depolarisation by the cloud material}

Any thermal material, like the warm, dense clouds inside the outer
cocoon prior to their dispersion, threaded by magnetic field will
contribute to the depolarisation of the radio synchrotron emission of
the cocoon material by Faraday rotation. In some FRII sources a clear
spatial association between the line emitting clouds and the
depolarisation is observed (e.g. Pedelty et
al. 1989\nocite{prms89}). As long as the clouds are still intact their
contribution to the depolarisation can not be distinguished from any
depolarisation occurring outside the radio cocoon. However, in DDRGs
and other large radio galaxies the cloud material should be spread out
over most of the volume of the cocoon and in this case the
depolarisation should show the characteristics of internal
depolarisation. Garrington \& Conway (1991)\nocite{gc91} find no
evidence for the presence of thermal material in the cocoons of a
sample of 47 FRII radio galaxies with a maximum linear size of 625
kpc. They give an upper limit of $5 \cdot 10^{-3}$ cm$^{-3}$ $\mu$G
for the product of the strength of the magnetic field along the line
of sight in the cocoon, $B_c$, and the number density of electrons,
$n_c$, in this region. From the model of KA we find the pressures in
the old cocoons of the DDRGs to be of order $\sim 10^{-13}$ erg
cm$^3$. Assuming equipartition between the energy density of the
magnetic field and that of the relativistic particles in the cocoon we
find $B \sim 3$ $\mu$G. The upper limit of Garrington \& Conway
(1991)\nocite{gc91} then corresponds to densities of roughly $2 \cdot
10^{-3}$ cm$^{-3}$ for the cocoon material. The densities within the
old cocoons of the DDRGs necessary to explain the properties of the
inner source structures are about three orders of magnitude lower than
this (see Tab. \ref{tab:res}). The internal depolarisation of the
radio emission of large radio galaxies caused by the material of the
dispersed warm clouds should therefore be unobservable. However, the
model presented here predicts that as more and more of the warm clouds
are shredded within the cocoons, the radio depolarisation observed
towards the cocoon of FRII radio sources should decrease with
increasing linear size. This effect may be masked by the additional
decrease in depolarisation caused by the fact that for large FRII
sources a larger fraction of the radio lobes is surrounded by material
of low density at large distances from the core which causes less
depolarisation than the material closer in (e.g. Strom \& J{\"a}gers
1988\nocite{sj88}). However, the shredding of warm clouds should lead
to a decrease in the number of `clumpy' regions of depolarisation,
possibly associated with optical line emission, in large FRIIs and
this difference may be detectable in high resolution depolarisation
maps of sources of different size.

\subsection{The large size of DDRGs}

For the cloud in the example presented in the previous section the
final dispersion takes place when the outer cocoon has a linear size
of 670 kpc.  If the shredding of line emitting clouds is the mechanism
which contaminates the cocoon of FRII radio sources then, because of
the long time it takes to disrupt these clouds, we expect to see large
aligned DDRGs only.  Indeed, all aligned DDRGs mentioned in paper I
have large linear sizes ($\ga 700$ kpc).  

If the jet flow in younger sources is disrupted without significantly
changing the direction of the jets, the new jets will pass through the
old cocoons at the bulk velocity of the jet material without
developing a strong jet shock because the warm clouds are still intact
and can not influence the dynamics of the new jets.  Those sources
will most likely follow a evolution much like that predicted by Clarke
\& Burns (1991), in which a restarted jet travels unhampered through
the old cocoon without a clear trace of its doing so.  Because of the
relativistic velocities of the jet material, the time it takes the new
jets to reach the edge of the old cocoon is about the light travel
time along the cocoon.  Since the old hotspots will fade extremely
fast after the energy input has stopped, such sources would only be
noticed during the short time their radio lobes do not show any hot
spots.

It is unclear how such a source will evolve further once the jet has
reached the old cocoon boundary again.  During the time that the jet
was off, the old cocoon will have continued to expand and it will
probably have a lower length-to-width ratio, since mainly the forward,
jet-driven motion will have slowed down and not the sideways
expansion.  The new hotspot, since it is driven by the restarted jet,
will advance much faster than the head of the old cocoon and it will
probably form a protrusion at the head of the old lobe (see also the
simulations of Clarke \& Burns 1991).  There are a number of radio
sources which clearly show protrusions in one or both radio lobes
(e.g. 3C\,79, Spangler, Myers \& Pogge 1984; 3C\,132, Neff, Roberts \&
Hutchings 1995).  These may therefore well be radio galaxies with
restarted jets.

\subsection{The velocity of the inner lobes}

Since the density in the old cocoon is lower than that of the
unperturbed IGM, it is expected that the inner structures advance much
faster than the outer structures would have at a similar distance from
the core.

Finding direct evidence for such high advance velocities, for instance
through spectral ageing studies, is difficult and often ambiguous (e.g
Alexander \& Leahy 1987).  If the inner structures indeed are the
result of a restarted jet, then the fact that we still detect the
outer radio lobes in the DDRGs already indicates that the inner lobes
must be advancing relatively fast.  Radio lobes which are not being
refuelled anymore are expected to fade away quickly, in a few times
$10^7$ yr, at most.  Since the radio powers of the outer structures
are quite normal for sources of their size (see paper I), the length
of time elapsed since the disconnection of the outer lobes from the
jet flow must be relatively short compared to their age.  Hence, the
inner structures must have grown relatively fast to their currently
observed size, certainly faster than the outer lobes would have
advanced at the time they had a similar size.

In the case of B\,1834+620 we have been able to constrain the advance
velocity and age of the inner structure (see paper III).  We find that
the velocity must lie within the range $0.2-0.3c$, depending only on
the orientation of the source.  This is much higher than what is
usually found in powerful radio galaxies ($0.01-0.1c$; e.g. Alexander
\& Leahy 1987, Scheuer 1995). The age is constrained to the range
between 2.6 and 5.8 Myr, which is in good agreement with the
prediction of the model presented in this paper (see
Tab. \ref{tab:res}).  Also, we have estimated an ambient density of
the inner lobes of $\sim 8\times10^{-7}$ cm$^{-3}$, three orders of
magnitude below what is generally found around the lobes of radio
galaxies (e.g. Alexander \& Leahy 1987), but in reasonable agreement
with the prediction from the model in this paper (see
Tab. \ref{tab:res}).

\section{Implications for the alignment effect}
\label{sec:opt}

The alignment of the UV and optical line emission with the radio axis
in FRII sources at high redshift, $z \ge 0.6$, is usually explained by
ionisation of the warm clouds by either the radiation from the AGN
(McCarthy 1993\nocite{pm93} and references therein) or by the emission
of gas shocked and heated by shocks (e.g. Dopita \& Sutherland
1995\nocite{ds95}). Aligned optical continuum emission is partly
caused by the nebular continuum emitted by the ionized clouds (Dickson
et al. 1995\nocite{dtscm95}) and may include contributions by
scattering of the AGN emission by the cloud material (Tadhunter et
al. 1987\nocite{tfbdr87}) and shock induced star formation (Rees
1989\nocite{mr89}, Begelman \& Cioffi 1989\nocite{bc89}). It is
interesting to note that all of the explanations mentioned above
postulate the existence of a two phase IGM; warm ($T_{cl} \sim 10^4$
K), dense clouds embedded in a hot ($T_x \sim 10^7$ K), less dense
background.

The model for the shredding of the warm, dense clouds presented here
is consistent with all of the explanations for the alignment
effect. The cores of the clouds may very well collapse after the
compression of the cloud by the bow shock while the outer cloud
regions are stable for a significant fraction of the life time of the
radio source. The material in this region can be ionized but will also
scatter the light of the AGN. The hot phase of the IGM may provide
some of the ionizing radiation after passing through the bow
shock. The following considerations show that the model is consistent
with the observed properties of the alignment effect.

We have shown in Sect. \ref{sec:disp} that the properties of the
warm, line-emitting clouds can change significantly when they pass
through the bow shock of an FRII source. Most optical emission from
FRII sources is observed in regions overlapping with the radio cocoon
which is the basis of the alignment effect. All physical quantities
derived from such observations for the warm clouds therefore apply to
their post-shock state within the cocoon of the FRII source. This has
important implications for their confinement in the hot IGM before
they are shocked.

To be stable the pre-shock cloud must be in pressure equilibrium with
the hot IGM. For the typical temperatures involved this implies
$n_{cl} / n_x \sim 10^3$, where $n_x$ is the density of the hot
IGM. The densities of these clouds derived from the observed line
emission coming from regions overlapping with the radio cocoon would
then imply much higher densities of the hot IGM than are derived from
X-ray observations of this material (e.g. Fabian et
al. 1987)\nocite{fcjt87}. However, because the clouds, for which these
observations have been made, may have been compressed by the bow shock
their pre-shock densities are much lower than those indicated by the
observations. For $M_b \sim 10$ we find that the pre-shock clouds are
stable within the density profile of the hot IGM in poor groups of
galaxies assumed above out to roughly 50 kpc from the centre of the
group. This explains the stability of the warm clouds in their hot
environment without the need to invoke additional contributions to the
thermal pressure of the hot IGM by cooling flows or other processes.

In some radio galaxies Ly$\alpha$ emission is detected well beyond the
radio cocoon but still aligned with the radio axis forming the so
called Ly$\alpha$ halo (Chambers, Miley \& van Breugel
1990\nocite{cmb90}, McCarthy et al. 1990\nocite{msdblde90}, Eales et
al. 1993\nocite{erdshl93}, van Ojik et al. 1996\nocite{orcmbm96},
Pentericci et al. 1997\nocite{prmcm97}). The source of ionizing
photons for the Ly$\alpha$ emission cannot be due to recently formed
stars and must be caused by the illumination of warm material by the
obscured AGN or by the shocked hot phase of the IGM (e.g. Meisenheimer
\& Hippelein 1992\nocite{mh92}). However, the warm clouds causing the
observed emission of the Ly$\alpha$ halos must still be in their
pre-shock state. This can explain observed differences in velocity
spread and ionisation parameter between the material in the halo and
that of the region of the cocoon.

Velocities of the warm clouds in regions overlapping with the cocoon
and in the more extended halo can be inferred from their
Doppler-broadened emission lines. In the Ly$\alpha$ halos typical
velocities are of order 500 km s$^{-1}$ which is comparable to the
sound speeds within the hot IGM and may indicate some large scale
movement like rotation (van Ojik et al. 1996\nocite{orcmbm96}). For
the line-emitting regions within the radio cocoons velocities of order
1000 km s$^{-1}$ are measured (McCarthy 1993\nocite{pm93} and
references therein). These somewhat higher velocities of the clouds
within the cocoon may be caused by the acceleration of the clouds by
the passage of the bow shock and the continuing momentum transfer from
the material in the cocoon to the clouds. The velocities within the
cocoon should be similar to the advance speed of the hot spots (Norman
et al. 1982\nocite{nsws82}). These are typically a few times 1000 km
s$^{-1}$ which agrees well with the observations. This may indicate
that the clouds observed in regions overlapping with the cocoon are
indeed located within the cocoon.

Furthermore the ratio of ionizing photons arriving at a warm cloud to
the number density of electrons available to absorb these photons
within the cloud, the ionisation parameter, $U$, will also change when
the cloud passes through the bow shock (e.g. Lacy et al.
1998\nocite{lrbr98}). The warm clouds in the cocoon will have been
compressed by the bow shock but the ionizing flux from the core is
roughly the same for clouds within the cocoon and those in front of
it. This will lead to a smaller value of $U$ for the clouds within the
cocoon compared to those in front of it. This can in principle be
tested observationally using line ratios like $[$O {\sc iii}$]$5007 / $[$O {\sc ii}$]$3727.

All models of the dynamical evolution of the radio cocoons of FRII
sources predict a correlation of linear size with age of the
source. The exact age of a source of a given linear size depends of
course on the power of its jets and also on the density of the
surrounding material. However, larger sources will in general be older
than smaller sources. Once the line-emitting clouds within the cocoon
are completely disrupted, the resulting temperatures (a few $10^8$ K)
and densities in this region imply a drastic decrease in the line
emission of these clouds. In the picture of the slow dispersion of the
line-emitting clouds within the radio cocoon sketched above we
therefore expect the strongest aligned emission in small FRII sources
while in larger sources the effect should be weaker and may vanish in
large sources. The same is true for the aligned optical and UV
continuum emission because the nebular continuum emission (Dickson et
al. 1995\nocite{dtscm95}) of the warm clouds will decrease as the
cloud material starts to be spread out over the volume of the
cocoon. If the cores of the warm clouds collapse and form stars, their
emission will also fade because of the aging of the stellar
population. Best, Longair \& R{\"o}ttgering (1996)\nocite{blr96} find
in a sample of 8 FRII radio galaxies, all at a redshift $z \sim 1$ and
of similar radio luminosity, that the alignment effect depends on the
linear size and therefore presumably on the age of the radio
source. Sources with larger linear sizes show a weaker alignment
effect than smaller objects as predicted by our model.

\section{Conclusions}

Based on the observed properties of aligned DDRGs we have developed a
model for their evolution. The symmetry of the inner source structure
strongly suggests that the interruption of the jet flow takes place in
the central AGN. One possible physical process is the infall of large masses
of gas onto the accretion disk, possibly caused by a (repeated) interaction with
a companion galaxy (see paper I). This scenario may explain the existence of
so-called X-shaped or winged FRII radio galaxies, if similar processes
have taken place in these sources and the angular momentum vector of
the infalling gas is very different to that of the pre-existing
accretion disk. These objects may then constitute the `misaligned'
DDRGs.

We extended the model of the radio luminosity evolution of FRII
sources of KDA to allow for sources in which the jets have stopped
supplying the cocoon with energy. We used this model to investigate
the evolution of the outer and the inner source structure. This
analysis suggests that the density in the outer cocoon created by the
old jets is insufficient to explain the observed properties of the
inner source structure. We argue that the most likely contamination of
the old cocoon with additional gas is the slow dispersion by the
effects of the bow shock surrounding the old cocoon of warm, dense
clouds embedded in the otherwise hot IGM. The observation of optical
and UV emission aligned with the radio axis in many FRII sources at
high redshift implies that the clouds must survive for a long time in
the cocoon. They are, however, eventually destroyed and provide the
environment for inner source structures in DDRGs. The long survival
times for these clouds derived here are consistent with the
observation that the currently known aligned DDRGs are all large
($\ge$ 700 kpc).  Also, it is consistent with the observation that in
$z\!\sim\!1$ 3CR radio galaxies the strength of the alignment effect
anti-correlates with the linear size of the radio sources. The lower
limits for the volume filling factors for the warm, dense clouds
derived from the density requirements of the inner source structures
are in good agreement with those obtained from optical observations.

The problem of the confinement of the warm clouds within the hot IGM
previously pointed out by various authors (e.g. Fabian et
al. 1987\nocite{fcjt87}) can be resolved by taking into account the
effects of the bow shocks on these clouds. The properties of the
clouds are derived from optical observations of clouds which may
already be located inside the radio cocoon. This implies that they
have been compressed by the bow shock leading to an increased density
and pressure for the cloud material compared to pre-shock clouds.

Polarization measurements of the radio emission from the cocoons of
FRII sources indicate a spatial association of peaks in the
depolarisation and concentrations of line-emitting material. Once the
cloud material has been spread out over the volume of the radio
cocoon, its depolarisation signature will be too weak to be detected.

The available data on the currently known DDRGs have not allow us to
put strong constraints on the model. Only in the case of the source
B\,1834+620 have we been able to limit the age and to estimate the
ambient density of the inner source. These values are in rough
agreement with the predictions from the model.

Many of the physical quantities we have derived for the DDRGs
discussed here are model dependent. The shape of the density
distribution within the region of the old cocoon in particular is
unknown. Moreover, we have assumed a density profile for the
environment of the outer source structures which may be typical for
low redshift FRII sources but in the absence of X-ray data we do not
have the means to test the validity of this assumption. The numerical
values of the derived source parameters should therefore be treated
with caution. However, we arrived at the main conclusions of this
paper, namely that the cocoons of aligned DDRGs and possibly other
FRII sources contain material additional to that supplied by their
jets, using only limiting cases of the models. This material is most
likely the remains of shredded line-emitting clouds. In view of the
observational evidence we believe that the contamination of the
cocoons of FRII sources by warm, dense clouds located in the IGM is an
important process in the evolution of these objects.

\section{Acknowledgments}

The authors would like to thank P. N. Best, A. G. de Bruyn, M. Lacy,
H. van der Laan and M. D. Lehnert for many stimulating discussions and
suggestions. We also thank the referee, J. P. Leahy, for his comments
on the manuscript. This work was supported in part by the Formation
and Evolution of Galaxies network set up by the European Commission
under contract ERB FMRX-- CT96--086 of its TMR programme.

\bibliography{crk} 
\bibliographystyle{mnras}

\end{document}